\documentclass[12pt]{article}

\usepackage{amsmath}
\usepackage{amsfonts}
\usepackage{amssymb}
\usepackage{latexsym}
\usepackage{graphicx}
\usepackage[english]{babel}
\input epsf.sty

\numberwithin{equation}{section} 


\newcommand{\be}{\begin{equation}}
\newcommand{\ee}{\end{equation}}
\newcommand{\ba}{\begin{array}}
\newcommand{\ea}{\end{array}}
\newcommand{\bqa}{\begin{eqnarray}}
\newcommand{\eqa}{\end{eqnarray}}
\begin{document}

\title{
Noncommutative Geometry Modified Non-Gaussianities of Cosmological
Perturbation }
 \vspace{3mm}
\author{{Kejie Fang$^{1}$, Bin Chen$^{1,2}$, Wei Xue$^{1}$}\\
{\small $^{1}$ Department of Physics, Peking University, Beijing 100871, P.R.China}\\
{\small $^{2}$ the Abdus Salam International Center for Theoretical Physics,}\\
{\small Strada Costiera 11, 34014 Trieste, Italy }}
\date{}
\maketitle

\begin{abstract}
We investigate the noncommutative effect on the non-Gaussianities of
primordial cosmological perturbation. In the lowest order of string
length and slow-roll parameter, we find that in the models with
small speed of sound the noncommutative modifications could
be observable if assuming a relatively low string scale. %in
%certain configuration of string length $l_s$ and the speed of
%sound on which the reliance is very sensitive.
In particular, the dominant modification of non-Gaussianity
estimator $f_{NL}$ could reach $O(1)$ in DBI inflation and
K-inflation. The corrections are sensitive to the speed of sound and
the choice of string length scale. Moreover the shapes of the
corrected non-Gaussianities are distinct from that of ordinary ones.
\end{abstract}

\section{Introduction}

Inflation\cite{inflation} is a very successful paradigm of the very
early universe. It can naturally solve several very tough
cosmological problems without fine tuning. Furthermore, it predicts
a nearly scale invariant Gaussian CMB spectrum, which has been
confirmed in the experiments\cite{WMAP3yr}. However, one of the
problems with inflation is that there are too many inflationary
models. It is necessary to find the signatures which could distinct
various models. With the development of precise cosmology, we expect
new experiments to constrain a large amount of inflation models and
to make the paradigm more clear. The scalar spectral index and its
running, the gravitational wave and non-Gaussian component of the
primordial fluctuations are among the important probes to detect
different inflation models. In this paper, we mainly discuss
non-Gaussianity.

In recent years, the non-Gaussianities of primordial perturbation
opens another window other than power spectrum to study different
scenarios of inflation models. After systematic
analysis\cite{Maldacena,ABMR,Seery,Seery2}, it is found that
non-Gaussianity estimator $f_{NL}$ is of order $O(\epsilon)$ in the
ordinary ``slow-roll" inflation and thus is unobservable. In some
particular models with non-trivial dynamics, such as ghost
inflation\cite{Ghost} and those with sufficiently small speed of
sound, including DBI model\cite{DBI,DBI in sky} and
K-inflation\cite{muk,muk1}, the unsuppressed non-Gaussianity is
potentially detectable in future experiments\cite{Shiu,DBI in sky}.
Furthermore, since the shape of non-Gaussianity is more multiple
than that of power spectrum, it could be used to distinguish
different inflation models through the classification of those
configuration of $k$ modes that determine the maximum of three-point
function\cite{shape}.

Non-Gaussianity could also be used to study various
trans-Planckian physics proposals. Deviation from the standard
Bunch-Davies vacuum was considered in \cite{Shiu,DBI in sky} where
the correction is found to be $O(\frac{H}{\Lambda})$ where
$\Lambda$ is the energy scale on which the modes are generated. In
\cite{high devi}, higher dimensional operators were introduced in
the inflaton Lagrangian. This modification could enhance the
non-Gaussian effect if the energy cutoff is not too high; but it
is difficult to exceed $f_{NL}\sim 1$. In this paper, we consider
another trans-Planckian scenario which is based on noncommutative
geometry.

In string theory, a promising candidate of quantum gravity, one
may use perturbative string or  nonperturbative object D-brane to
probe the spacetime geometry. Due to the extensive nature of the
string, or stringy effect, or strong string interaction, the
picture of spacetime geometry in string theory could be very
different from the usual one. Especially, very near the
cosmological singularity, the usual concept of commutative
geometry may break down completely. A better description could be
noncommutative geometry\cite{Connes}, in terms of the algebra
generated by noncommutative coordinates \be [x^\mu,
x^\nu]=i\theta^{\mu\nu}\ .\ee One natural way to get
noncommutative geometry is to consider the D-brane in the presence
of constant NS-NS magnetic B field\cite{witten99}. In this case,
the spatial coordinates are noncommutative \be [x^i,
x^j]=i\theta^{ij}\ ,\label{nc}\ee where $\theta^{ij}$ depends on
the background flux, while space-time is commutative. Similarly
one can obtain space-time noncommutativity by placing D brane in
constant electric field, but the theory is no more unitary in this
case\cite{tsnunitary1,tsnunitary2,tsnunitary3,tsnunitary4,tsnunitary5}.
%Actually, we do not have a clear understanding of the space-time
%noncommutativity to date, which conflicts with our current
%understanding of quantum mechanics, although this is technically
%achievable.

Another way to realize the noncommutative relation among
coordinates, especially between space and time, is to start from the
stringy uncertainty relation \be \Delta x_p \Delta t\geq l_s^2
\label{SUP} .\ee Equivalently one may assume $[t,x_p]=il^2_s$. This
kind of noncommutative relation have been applied to the study of
inflationary universe\cite{branden,Huang,Tsujikawa:2003gh}. In
\cite{branden}, Brandenberger and Ho started from (\ref{SUP}) and
discussed the cosmological implications of such relation. To keep
the background isometry intact, they actually considered the
noncommutative relation between radial coordinate $r$ and time $t$,
however, as we mentioned before, the space-time noncommutative
relation may violate the unitarity of the theory.

In this paper, in order to keep unitarity, we choose to set the
space-time components $\theta^{0i}$ zero. The cosmological imprint
based on this kind of noncommutativity has been studied by many
authors\cite{cgs,egks,egks2,ms,noncommutative}. Meanwhile, without
losing generality, we choose a particular frame in which the only
non-vanishing space-space component of $\theta^{\mu\nu}$ in
comoving coordinate is \be [x^1,x^2]=i\theta^{12}.\label{nc}\ee We
find the calculation is much more involved if one chooses to keep
all possible spatial noncommutativity but this won't change the
main result we obtained from this particular frame. We also
require that noncommutativity following from (\ref{nc}) only
dominates at small length scale. % as a correspondence principle.
A simple form of $\theta^{12}$ realizes this condition
is\cite{noncommutative} \be
\label{nc1}\theta^{12}=\frac{l_s^2}{a^2},\ee where $l_s$ is the
string length. In this case, the noncommutativity in physical
coordinate remains to be a constant, $[x_p^1, x_p^2]=il_s^2$, and
the noncommutativity of comoving coordinate is diluted by the
expansion of universe. However, with this choice, the isotropy of
the spacetime is now broken. Nevertheless, if the noncommutativity
is very small, the breaking of isotropy could be ignorable but the
physical implication could be observable. In this paper, we will
focus on this case.
%Liu:2004xg,Cai:2004ur,
%Kim:2004,Calcagni,Bamba:2004cu,Zhang:2006ta,Cai:2007xr,xue}.}

In most of the inflation models, the non-Gaussianity is quite small
and hard to be detected. The noncommutative corrections of
non-Gaussianity is even smaller. This drives us to work on the
models with big non-Gaussianity. We will discuss the noncommutative
effect in string-inspired DBI inflation model and k-inflation model.
We find that in both models the noncommutative modification of
non-Gaussianity estimator can reach $O(1)$ or even larger with a
small speed of sound and a relatively low string scale. We also
determine the shape of corrected non-Gaussianity, which is different
from the usual ones. Though the non-Gaussianity estimator could be
over-shadowed, the shape is a distinctive signature to be probed by
the future experiments. On the contrary, future experiments could
set bound of the noncommutativity and be used for crosschecking
along with other experiments, like atomic experiments.

The paper is organized as follows. In section 2, we give a brief
review of the ADM formalism to discuss the perturbations in a
general inflation lagrangian. In section 3, we study the
noncommutative modifications to general inflation models. In section
4, we focus on two inflation models with observable non-Gaussianity.
In section 5, we end with conclusion and discussions.

%In the period of inflation, the Hubble scale is nearly GUT scale
%or string scale. In this scale, the quantum effect of gravity,
%plays a great part in describing inflation. So we expect
%inflation can be used to test the physics of quantum gravity with
%the scale which cannot be attained in the earth experiments. In
%string theory, the appearance of spacetime noncommutativity is
%natural. The proper space and time should satisfy the following
%relation \be \Delta x_p \Delta t\geq l_s^2 \ .\ee There are some
%different methods to realize spacetime noncommutativity in
%inflation\cite{branden,noncommutative}. The recent discussions of
%noncommutative inflation can be found in. Brandenberger and Ho
%\cite{branden} give an ansatz of spacetime noncommutativity, in
%which the spectrum runs from blue to red. And in
%\cite{noncommutative}, which is our starting point in this paper,
%it is suggested that spacetime noncommutativity for comoving
%coordinate $x^\mu$ in curved spacetime to be defined as

\section{Perturbations in general inflation models}

Let us start with a general Lagrangian $P(X,\phi)$ which can be
used to describe a broad class of inflation models. The action is
of the form as follows,
\begin{equation}\label{action}
S=\frac{1}{2}\int d^{4}x\sqrt{-g}[R+2P(X,\phi)],
\end{equation}
where $\phi$ is the inflaton field and
$X=-\frac{1}{2}g^{\mu\nu}\partial_{\mu}\phi\partial_{\nu}\phi$. We
set Planck mass $M_{pl}=(8\pi G)^{-\frac{1}{2}}=1$ and adopt the
metric signature $(-1,1,1,1)$.

In order to proceed, it is convenient to work in the ADM metric
formalism,
\begin{equation}
d^{2}s=-N^{2}d^{2}t+h_{ij}(dx^{i}+N^{i}dt)(dx^{j}+N^{j}dt),
\end{equation}
where $h_{ij}=a^{2}(1+2\zeta)\delta_{ij}$, with $a$ the scale factor
which grows quasi-exponentially during inflation and $\zeta$ the
scalar perturbation of metric. In this paper, we do not consider the
tensor perturbation. Since in the ADM formalism $N$ and $N^{i}$ are
Lagrangian multipliers, the metric in terms of $\zeta$ can be
determined by solving the constraint equations of $N$ and $N^{i}$ in
a given gauge, rather by solving the Einstein equation, which makes
the calculation relatively simpler.

There are two main gauges in which we can calculate the
perturbation action. One gauge is the comoving gauge, that is
$\delta\phi=0$. In this gauge, $\zeta$ is the curvature scalar on
a comoving hypersurface and it directly seeds on the later
generation of large scale structure and anisotropy of microwave
background radiation. It is straightforward to find out that
$\zeta$ remains constant after horizon exit. Although the physical
meaning is manifest within this gauge, it is rather complicated to
analyze the order of the perturbation action with respect to the
slow-roll parameters, which is important in determining the
magnitude of correlation functions. In this gauge, the third order
action of perturbation is apparently of order $O(\epsilon^0)$,
where $\epsilon$ represents the slow-roll parameter; however after
doing a lot of integration by part the action is actually of order
$O(\epsilon^2)$\cite{Maldacena,Shiu}. The other gauge with
$\zeta=0$ is called uniform density gauge. The gauge
transformation linking the two gauge is
$\zeta=\frac{H}{\dot{\phi}}\delta\phi$, where $H$ is the Hubble
parameter $H=\frac{\dot a}{a}$. One could easily recover the exact
order $O(\epsilon^{2})$ of the action by doing a gauge
transformation of the perturbation action of $\delta\phi$
\cite{Maldacena}. In order to calculate the correlation function
of metric perturbation in this gauge, we first calculate the
correlation function of $\delta\phi$ and then transform it to
$\zeta$ just after horizon exit, which remains constant ever
since.

 Below we adopt the uniform density gauge
$\zeta=0$ to carry out the calculation. Substituting the ADM
metric into the action(\ref{action}), we get
\begin{equation}\label{ADMaction}
S=\frac{1}{2}\int dtd^{3}x\sqrt{h}N(R^{(3)}+2P)+\frac{1}{2}\int
dtd^{3}x\sqrt{h}N^{-1}(E_{ij}E^{ij}-E^{2}),
\end{equation}
where $h=\det(h_{ij})$,
$E_{ij}=\frac{1}{2}(\dot{h}_{ij}-\nabla_{i}N_{j}-\nabla_{j}N_{i})$,
$E=E_{i}^{i}$. $R^{(3)}$ is the Ricci scalar calculated in the
three-dimensional hypersurface with metric $h_{ij}$, and
$\nabla_{i}$ is the covariant differential coefficient defined on
the hypersurface. Since $N$ and $N^{i}$ are Lagrangian
multipliers, we can obtain two constraint equations from them,
which are
\begin{equation}\label{N}
R^{(3)}+2P-2N^{-2}P_{,X}(\dot{\phi}^{2}-2N^{i}\dot{\phi}\partial_{i}\phi+N^{i}N^{i}\partial_{i}\phi\partial_{j}\phi)
-N^{-2}(E_{ij}E^{ij}-E^{2})=0,
\end{equation}
\begin{equation}\label{Ni}
-NP_{,X}(2N^{-2}\dot{\phi}\partial\phi-\frac{2N^{j}}{N^{2}}\partial_{i}\phi\partial_{j}\phi)+2(\nabla_{j}(N^{-1}E^{j}_{i})
-\nabla_{i}(N^{-1}E))=0.
\end{equation}

We divide $\phi$ into the isotropic background $\phi_{0}(t)$ and
the fluctuation $\varphi$, $\phi=\phi_{0}+\varphi$. In order to
evaluate the third order action of perturbation, we only need to
solve the equations of $N$ and $N^{i}$ to the first order of
$\varphi$, since the second and third order of solutions, when
substituted into the action, will be multiplied with the first and
zeroth order Hamiltonian constraint of the action respectively and
thus vanish. In fact, to calculate the $n$-th order action of
perturbation, one only needs the solutions of $N$ and $N^{i}$ to
$(n-2)$-th order \cite{Shiu}. Following \cite{Maldacena}, we
decompose $N^{i}$ into two parts
$N_{i}=\tilde{N_{i}}+\partial_{i}{\psi}$, where
$\partial_{i}{\tilde N^{i}}=0$ and $N_{i}$ is lowered by
${h_{ij}}$ through $N^i$. Then we expand them to the first order
of $\varphi$,
\begin{equation}\label{expansion}
N=1+\alpha_{1},\quad \tilde N_{i}=N_{i}^{(1)},\quad \psi=\psi_{1},
\end{equation}
where $\alpha_{1}, N_{i}^{(1)}$ and $\psi_{1}$ are of order
$O(\varphi)$. Substituting (\ref{expansion}) into (\ref{N}) and
(\ref{Ni}), and solving them to $O(\varphi)$, we obtain the
solution,
\begin{equation}\label{alpha}
\alpha_{1}=\frac{P_{,X_0}\dot{\phi_{0}}}{2H}\varphi, \quad
N_{i}^{(1)}=0,
\end{equation}
\begin{eqnarray}\label{psi}&&
\partial^{2}\psi_{1}=\frac{1}{4H}\big((2P_{,\phi_0}+\dot{\phi_{0}^3}H^{-1}P^{2}_{,X_0}-2P_{,X_0\phi_0}\dot{\phi_{0}^{2}}+\dot{\phi_{0}^{5}}H^{-1}P_{,X_0}P_{,X_0X_0}
\varphi\nonumber\\
&&\qquad\quad-6HP_{,X_0}\dot{\phi_{0}})\varphi-(2P_{,X_0}\dot{\phi_{0}}+2P_{,X_0X_0}\dot{\phi_{0}})\dot{\varphi}\big),
\end{eqnarray}
where the subindex $0$ represents the background value.

It will be more succinct to express the solutions using the
slow-roll parameter $\epsilon$ and the ``speed of sound'' $c_s$,
\begin{equation}
\epsilon=-\frac{\dot{H}}{H^2}=\frac{X_0P_{,X_0}}{H^2M_{pl}^2},
\end{equation}
\begin{equation}
c_s^2=\frac{dP}{dE}=\frac{P_{,X_0}}{P_{,X_0}+2X_0P_{,X_0X_0}},
\end{equation}
where $E=2XP_{,X}-P$ is the energy of inflaton field. Keeping with
the lowest order of slow-roll parameter, equations
(\ref{alpha})(\ref{psi}) can be written as
\begin{equation}
\alpha_1=\frac{H\epsilon}{\dot\phi_0}\varphi,\label{alpha1}
\end{equation}
\begin{equation}
\partial^{2}\psi_{1}=(\frac{P_{,\phi_0}}{2H}-\frac{3H^2}{\dot\phi_0}\epsilon)\varphi-\frac{H\epsilon}{\dot\phi_0c_s^2}\dot\varphi.\label{psi1}
\end{equation}
We omit the subindex $1$ in these expressions in the following
section for simplicity.

Substituting (\ref{alpha1}) and (\ref{psi1}) into
(\ref{ADMaction}) and expanding it to the second order of
perturbation, we attain the free field action of fluctuation
\begin{equation}
S_2=\frac{1}{2}\int
d^4xa^3(\frac{P_{,X_0}}{c_s^2}\dot\varphi^2-P_{,X_0}(\partial\varphi)^2).
\end{equation}

The equation of motion is
\begin{equation}\label{motioneqn}
(a\varphi)^{''}+(c_s^2k^2-\frac{a^{''}}{a})a\varphi=0,
\end{equation}
where the prime denotes derivative with respect to conformal time
and we have assumed $P_{,X_0}$ and $c_s$ to be time independent.
The classical solution is
\begin{equation}
\varphi=\frac{\dot\phi_0}{\sqrt{4\epsilon
c_sk^3}}(1+ic_sk\eta)e^{-ic_sk\eta},
\end{equation}
where we choose the standard Bunch-Davies vacuum to fix the
coefficient. With the same procedure one could obtain the third
order action of perturbation which has been studied thoroughly in
\cite{Shiu} in the comoving gauge.

\section{Noncommutative modification}

In the noncommutative spacetime,  the functions are better described
by the operators in Hilbert space. This is very similar to the case
in quantum mechanics, where one has a noncommutative phase space.
The product of functions may be taken as the multiplication of the
operators. But an efficient way to define product is by so-called
Moyal product, whose expansion in curved spacetime
gives\cite{noncommutative}
 \be f \star g \equiv
\sum_{k=0}^{\infty} \frac{1}{k!} \left( \frac{i}{2} \right)^k
\theta^{\mu_1 \nu_1} \cdots \theta^{\mu_k \nu_k} \left( D_{\mu_1}
\cdots D_{\mu_k} f \right) \left( D_{\nu_1} \cdots D_{\nu_k} g
\right) , \label{star} \ee where $D_{\mu}$ is the covariant
derivative coefficient in the curved spacetime.

The star product could be extended to multiple function situation,
at quadratic order in $\theta^{\mu\nu}$, \be f_1\star \cdots \star
f_n=
(1+\frac{i}{2}\theta^{\mu\nu}\sum_{a<b}D_\mu^aD_\nu^b-\frac{1}{8}\theta^{\mu\nu}\theta^{\rho\sigma}\sum_{a<b,c<d}D_\mu^aD_\nu^bD_\rho^cD_\sigma^d)f_1\cdots
f_n.\ee

To incorporate the noncommutative effect, we replace the ordinary
product in the inflaton Lagrangian $P(X,\phi)$ with the star
product. Suppose that $P(X,\phi)$ relies on $\phi$ through
function $V(\phi)$ \footnote{If there are more than one such
functions, one just needs to include all the corrections from each
function in (\ref{deltaS}) .}, which could be the inflaton
potential or the warping factor in DBI model. Without losing
generality, we simply consider $V(\phi)$ of the form
$V(\phi)=\phi^{n},n\ge 1$, since one could always do Taylor's
expansion of a generic function. When apply the star product in
the generic Lagrangian, noncommutativity changes the forms of
dynamic term $X=-\frac{1}{2}\partial_{\mu}\phi\partial^{\mu}\phi$
and scalar function $V(\phi)$ as follows
\bqa X&=&-\frac{1}{2}(\partial_\mu \phi)\star(\partial^\mu \phi)\nonumber\\
            &=&-\frac{1}{2}(\partial_\mu \phi)(\partial^\mu
            \phi)+\frac{1}{8}(\theta^{12})^2g^{\mu\nu}(D_1D_1D
_\mu\phi D_2D_2D_\nu\phi\nonumber\\
            &&- D_1D_2D _\mu\phi
D_2D_1D_\nu\phi)+O(\theta^3) ,\eqa and \bqa V&=& \phi \star \cdots
\star \phi \nonumber\\&=&
\phi^n-\frac{n(n-1)}{8}(\theta^{12})^2\phi^{n-2}(D_1D_1\phi
D_2D_2\phi-D_1D_2\phi D_2D_1\phi)\nonumber\\
&&-\frac{n(n-1)(n-2)}{24}(\theta^{12})^2\phi^{n-3}(D_1D_1\phi
D_2\phi
D_2\phi+D_2D_2\phi D_1\phi D_1\phi\nonumber\\
&&-D_1D_2\phi D_1\phi D_2\phi-D_2D_1\phi D_2\phi D_1\phi)\nonumber\\
&&+O(\theta^3). \eqa

We find that the lowest order noncommutative modification term is of
order $O(\theta^{2})$. We only consider the correction of the lowest
order $\theta^2$ in this paper, and denote them as $\delta_\theta X$
and $\delta_\theta V$ respectively, \bqa \delta_\theta
X&=&\frac{1}{8}(\theta^{12})^2g^{\mu\nu}(D_1D_1D _\mu\phi
D_2D_2D_\nu\phi- D_1D_2D _\mu\phi D_2D_1D_\nu\phi), \eqa \bqa
\delta_\theta V &=&
-\frac{n(n-1)}{8}(\theta^{12})^2\phi^{n-2}(D_1D_1\phi
D_2D_2\phi-D_1D_2\phi D_2D_1\phi)\nonumber\\
&&-\frac{n(n-1)(n-2)}{24}(\theta^{12})^2\phi^{n-3}(D_1D_1\phi
D_2\phi
D_2\phi+D_2D_2\phi D_1\phi D_1\phi\nonumber\\
&&-D_1D_2\phi D_1\phi D_2\phi-D_2D_1\phi D_2\phi D_1\phi). \eqa

One should keep in mind that all the covariant derivatives are
calculated within the ADM formalism, which makes the evaluation much
more involved. In some other papers in calculating large
unsuppressed non-Gaussianity \cite{Ghost,DBI,high devi}, because of
the particularity of the actions (they could themselves, by doing
serial expansion, generate cubic terms of perturbation without
involving metric corrections, i.e. $\alpha$, $\psi$), the authors
ignore the correlation between metric correction and inflaton
perturbation in the uniform density gauge, which is subleading in
slow-roll parameter, and do the calculation with isotropic FRW
metric. Our calculation turns out to be the same situation when
considering the noncommutativity correction.

The change of the inflation action is \be\delta_\theta S = \int
d^{4}x\sqrt{-g}(P_{,X}\delta_\theta X+P_{,V}\delta_\theta V).\ee To
obtain the third order action of perturbation, one needs to serially
expand $P_{,X}$ and $P_{,V}$ around the background value and
multiply with the corresponding terms in $\delta_\theta X$ and
$\delta_\theta V$ which generate overall cubic terms of
perturbation.

To simplify the calculation, we need to pick out the terms of
leading order of slow-roll parameter. We find in this case, the
leading order is $O(\epsilon)$ for $\zeta$ after gauge
transformation. We decompose $\delta_\theta X$ into terms of
different order of perturbation,
\begin{equation}
\delta_\theta X=(\delta_\theta X)_3+(\delta_\theta
X)_2+(\delta_\theta X)_1+(\delta_\theta X)_0,
\end{equation}
where $(\delta_\theta X)_3$ represents the cubic terms, etc., and we
do not consider the higher order of perturbation since we are only
going to calculate the three-point function. The terms of lowest
order of slow-roll parameter in $(\delta_\theta X)_3$ are those
composed of product of two inflaton perturbation $\varphi$ and one
metric correction, i.e. $\alpha$ and $\psi$. However, we do not need
to take them into account to obtain the final third order action of
perturbation by the reason that they will generate terms of second
order in slow-roll parameter after doing the gauge transformation to
$\zeta$, which turns out to be the subleading terms in our
calculation. As for $(\delta_\theta X)_0$ which is composed of terms
of $\dot\phi_0$, it also results in subleading order terms when
multiplied with the serially expanding terms of $P_{,X}$. In short,
we only need $(\delta_\theta X)_2$ and $(\delta_\theta X)_1$.

The terms of lowest order of slow-roll parameter in
$(\delta_\theta X)_2$ are the products of two inflaton
perturbation, which are summarized as follows,
\begin{eqnarray}
(\delta_\theta
X)_2&=&\frac{1}{8}(\theta^{12})^2(\frac{1}{a^2}\partial_1^2\partial_i\varphi\partial_2^2\partial_i\varphi-\frac{1}{a^2}\partial_1\partial_2\partial_i\varphi\partial_1\partial_2\partial_i\varphi
-H\partial_1^2\partial_i\varphi\partial_i\dot\varphi\nonumber\\
&& -H\partial_2^2\partial_i\varphi\partial_i\dot\varphi
+H^2\partial_1^2\partial_i\varphi\partial_i\varphi+H^2\partial_2^2\partial_i\varphi\partial_i\varphi+\dot
a^2\partial_1\dot\varphi\partial_1\dot\varphi\nonumber\\
&& +\dot a^2\partial_2\dot\varphi\partial_2\dot\varphi+\dot
a^2\partial_i\dot\varphi\partial_i\dot\varphi -2H\dot
a^2\partial_1\dot\varphi\partial_1\varphi-2H\dot
a^2\partial_2\dot\varphi\partial_2\varphi  \nonumber\\
&& -2H\dot a^2\partial_i\dot\varphi\partial_i\varphi+(H\dot
a)^2\partial_1\varphi\partial_1\varphi+(H\dot
a)^2\partial_2\varphi\partial_2\varphi\nonumber\\
&& +(H\dot
a)^2\partial_i\varphi\partial_i\varphi-\partial_1^2\dot\varphi\partial_2^2\dot\varphi+\partial_1\partial_2\dot\varphi\partial_1\partial_2\dot\varphi
-(Ha\dot a)^2\dot\varphi^2 \nonumber\\
&& -Ha\dot a\dot\varphi\partial_1^2\dot\varphi+Ha\dot
a\dot\varphi\partial_2^2\dot\varphi+2H\partial_1^2\dot\varphi\partial_2^2\varphi+2H\partial_1^2\varphi\partial_2^2\dot\varphi\nonumber\\
&& + 2H^2a\dot a\dot\varphi\partial_1^2\varphi+2H^2a\dot
a\dot\varphi\partial_2^2\varphi-4H\partial_1\partial_2\dot\varphi\partial_1\partial_2\varphi
\nonumber\\
&&-4H^2\partial_1^2\varphi\partial_2^2\varphi+4H^2\partial_1\partial_2\varphi\partial_1\partial_2\varphi
+a\dot a {\partial_1}^2 \dot \varphi \ddot \varphi + a \dot a
{\partial_2}^2 \dot \varphi \ddot \varphi \nonumber\\
&&- a^2 {\dot a}^2 \ddot \varphi \ddot \varphi - {\dot a}^2 \ddot
\varphi ( 2 {\partial_1}^2 \varphi +2 {\partial_1}^2 \varphi- a \dot
a \dot \varphi)),
\end{eqnarray}
where $i$ should be summed from $1$ to $3$. And $(\delta_\theta
X)_1$ is
\begin{equation}
(\delta_\theta X)_1=-\frac{1}{8}(\theta^{12})^2Ha\dot
a\dot\phi_0(\partial_1^2\dot\varphi+\partial_2^2\dot\varphi-2H\partial_1^2\varphi-2H\partial_2^2\varphi+2Ha\dot
a\dot\varphi).
\end{equation}

Following the same procedure, we decompose $V$ as \be \delta_\theta
V=(\delta_\theta V)_3+(\delta_\theta V)_2+(\delta_\theta
V)_1+(\delta_\theta V)_0.\ee According to the same reason as that of
the case of $\delta_\theta X$, we can just consider the
\textit{leading} terms in $(\delta_\theta V)_2$ and $(\delta_\theta
V)_1$, which are
\begin{eqnarray}
(\delta_\theta
V)_2&=&-\frac{n(n-1)}{8}(\theta^{12})^2\phi_0^{n-2}(\partial_1^2\varphi\partial_2^2\varphi
-\partial_1\partial_2\varphi\partial_1\partial_2\varphi +(a\dot
a)^2\dot\varphi^2-\nonumber\\
&&a\dot a\dot\varphi\partial_1^2\varphi-a\dot
a\dot\varphi\partial_2^2\varphi),
\end{eqnarray}
\begin{equation}
(\delta_\theta
V)_1=-\frac{n(n-1)}{8}(\theta^{12})^2\phi_0^{n-2}(2(a\dot
a)^2\dot\phi_0\dot\varphi-a\dot
a\dot\phi_0\partial_1^2\varphi-a\dot
a\dot\phi_0\partial_2^2\varphi).
\end{equation}

The whole change of third order action of perturbation due to
noncommutative geometry in leading order of slow-roll parameter can
be written as
\begin{eqnarray}\label{deltaS}
\delta_\theta S_3 &=& \int d^{4}x\sqrt{h}(P_{,X_0X_0}(\delta_g
X)_1(\delta_\theta
 X)_2+P_{,X_0X_0}(\delta_g X)_2(\delta_\theta
 X)_1\nonumber\\
 &&\qquad\qquad+P_{,X_0\phi_0}\varphi(\delta_\theta
 X)_2+\frac{1}{2}P_{,X_0X_0X_0}(\delta_g X)_1^2(\delta_\theta
 X)_1\nonumber\\
 && \qquad\qquad+P_{,V_0X_0}(\delta_gX)_1(\delta_\theta V)_2+P_{,V_0X_0}(\delta_gX)_2(\delta_\theta V)_1),
\end{eqnarray}
where $\delta_g X=X-X_0$, and $(\delta_g
X)_1=\dot\phi_0\dot\varphi$, $(\delta_g
X)_2=\frac{1}{2}(\dot\varphi^2-(\partial\varphi)^2)$, where we have
picked out the terms with least $\dot\phi_0$ to reduce the order of
slow-roll parameter. Although we have picked out the leading order
terms, the result is still lengthy. We write the result of
(\ref{deltaS}) in Appendix A. To write the result, we define two new
parameters,
 \begin{eqnarray}
 &&c_V^2=\frac{P_{,X_0}}{P_{,X_0}+2V_0P_{,X_0V_0}}\label{c_v},\\
 &&\sigma=X_0X_0X_0P_{,X_0X_0X_0}.
 \end{eqnarray}
As we pointed out above, all the terms except those with $\sigma$
coefficient are of order $O(\epsilon)$ in $\zeta$ after a gauge
transformation $\varphi=\frac{\dot{\phi}}{H}\zeta$. These two
parameters vanish in the general ``slow-roll'' inflation models, but
could be non-trivial in some particular models. We will see in some
case, they determine the dominance of the correction of
non-Gaussianity.

There are some subtle differences between calculating the
modification of three-point function and two-point function due to
noncommutativity. In calculating two-point function, one has to
solve the equation of motion of perturbation which is in general
hard to solve with the presence of noncommutative coordinates(for
a solvable example, see \cite{noncommutative}). %We list the
%equation of motion with noncommutative geometry modification in
%Appendix A.
 In Ref.\cite{branden}, the author developed another
way to encode the noncommutative effect into the power spectrum
without solving the equation of motion.

However, even with a solvable equation, we do not need the modified
classical solution to calculate noncommutative correction of
three-point function. The three-point function is calculated through
\be \langle \zeta^3(t)
\rangle=-i\int_{t_0}^tdt'\langle[\zeta^3(t),H_{int}(t')]\rangle\ee
in tree level. So, to evaluate the modification of three-point
function, which is denoted by $\langle \zeta^3(t) \rangle_{\theta}$
below, in the lowest order of $\theta$, we divide the Poisson
bracket into two groups:  \be \langle \zeta^3(t)
\rangle_{\theta}=-i\int_{t_0}^tdt'\langle[\zeta_c^3(t),\delta_{\theta}H_{int}(t')]\rangle
-i\int_{t_0}^tdt'\langle[3\zeta_c^2(t)\delta_{\theta}\zeta(t),H_{int}(t')]\rangle,\ee
where $\zeta_c$ is the commutative solution. Since the primordial
Hamiltonian without noncommutative modification is of order
$\epsilon^2$ in slow-roll parameter, thus the second group
contributes terms subleading in slow-roll parameter, as we emphasize
the leading order of the modification part of Hamiltonian is
$O(\epsilon)$. While the order of $\theta$ are $\theta^2$ in both
case, so the leading modification is obtained from the first group
of Poisson bracket which is summarized in Appendix B.

Another point deserved mention is that the constraint equations of
$N$ and $N_i$, and thus the solutions (\ref{alpha1})(\ref{psi1})
obtain corrections of order $\theta^2$ in noncommutative coordinate.
However, the same as the above analysis, they contribute terms of
subleading order of $\theta$ or $\epsilon$ either in $\delta_\theta
S$ or $S$ and thus we do not need to consider them.

\section{Model testing}
In this section, we evaluate the effect of noncommutativity on the
non-Gaussianities of perturbation in some particular models. Note
all the background field and Hubble scale are estimated at horizon
crossing, namely at the time about $60$ e-foldings before ending of
inflation.

In the ordinary ``slow-roll" inflation with Lagrangian $P=X-V$,
since \be c_s=1,\quad c_V=1,\quad \sigma=0,\ee it turns out that
the modification is zero on the level of first order of slow-roll
parameter. Let us turn to the models with significant
non-Gaussianity.

\subsection{DBI model}

\begin{figure}%
\begin{center}%
\mbox{\epsfxsize=8cm\epsffile{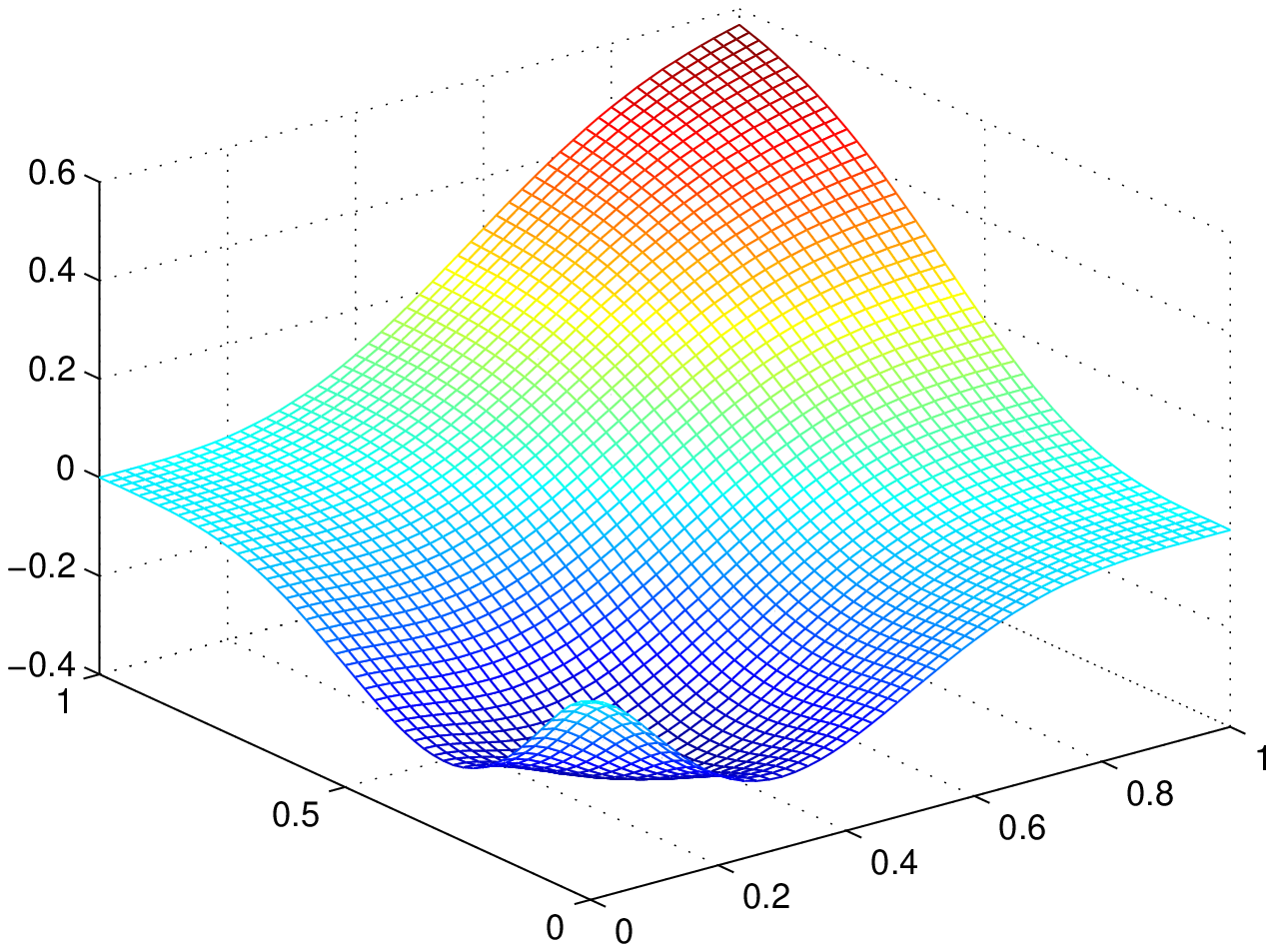}} \caption{\label{A1}The shape
of $\mathcal{A}_1/k_1k_2k_3$.}
\end{center}%
\end{figure}

DBI inflation \cite{DBI,DBI in sky,Chen1,Chen2} is motivated by
brane inflation scenario\cite{KKLMMT,Tye,DSS,BMNQRZ,ST} in warped
compactifications. The effective Lagrangian is
\begin{equation}
P(X,\phi)=-f(\phi)^{-1}\sqrt{1-2f(\phi)X}+f(\phi)^{-1}-V(\phi),
\end{equation}
where $f$ is the warp factor $f=\frac{\lambda}{\phi^4}$, and
$\lambda$ depends on flux number. The value of speed of sound
$c_s$, as well as other two parameters $c_f$ ($f$ substituting for
$V$ in (\ref{c_v}) represents warp factor $f(\phi)$) and $\sigma$
are as follows,
\begin{eqnarray}
&&c_s=\sqrt{1-\dot\phi^2f(\phi)},\\
&&c_f=\sqrt{1-\dot\phi^2f(\phi)},\\
&&\sigma=-\frac{3}{8}\dot\phi^6f^2(1-\dot\phi^2f)^{-\frac{5}{2}}.
\end{eqnarray}
We find $c_f$ coincides with $c_s$ in DBI model, and thus it becomes
as small as the speed of sound. The leading correction terms with
respect to $c_s$ comes from (\ref{three}) are \bqa \label{DBI}
\langle\zeta_{k_1}\zeta_{k_2}\zeta_{k_3}\rangle_\theta&=&
i(2\pi)^3\delta(\vec{k}_1+\vec{k}_2+\vec{k}_3)\frac{H^4l_s^4}{32\epsilon^2c_s^4M_{pl}^4}\frac{1}{\prod(2k_i^3)}[2i\frac{H^4}{c_s^4}
((k_1^a)^2(k_2^b)^2\nonumber\\
&&-k_1^ak_1^bk_2^ak_2^b)(\vec{k}_1\cdot\vec{k}_2)k_3^2(24\frac{1}{K^5}+120\frac{k_1+k_2}{K^6}+720\frac{k_1k_2}{K^7})\nonumber\\
&&+4\frac{iH^3}{c_f^2c_s^2}\frac{\phi_0}{\sqrt\lambda}((k_1^a)^2(k_2^b)^2-k_1^ak_1^bk_2^ak_2^b)(\vec{k}_1\cdot\vec{k}_2)(8\frac{1}{K^3}\nonumber\\
&&+24\frac{k_1k_2+k_2k_3+k_3k_1}{K^5}+120\frac{k_1k_2k_3}{K^6})\nonumber\\
&&-6iH^2\frac{1}{c_f^2}\frac{\phi_0^2}{\lambda}(((k_1^a)^2+(k_1^b)^2)k_2k_3(2\frac{1}{K}\nonumber\\
&&+2\frac{k_1k_2+k_2k_3+k_3k_1}{K^3}+6\frac{k_1k_2k_3}{K^4})+2((k_1^a)^2(k_2^b)^2\nonumber\\
&&-k_1^ak_1^bk_2^ak_2^b)k_3^2(2\frac{1}{K^3}+6\frac{k_1+k_2}{K^4}+24\frac{k_1k_2}{K^5}))\nonumber\\
&&-i\frac{4\sigma
c_s^2H^2}{M_{pl}^2\epsilon}((k_1^a)^2+(k_1^b)^2)(24k_1^2k_2^2k_3^2\frac{1}{K^5}\nonumber\\
&&-2k_2^2k_3^2(2\frac{1}{K^3}+6\frac{k_1}{K^4}))+perm.], \eqa where
$a$ and $b$ denote the first and second component of $k$ vector
respectively, $K=k_1+k_2+k_3$ and $perm.$ denotes all the other
terms obtained by rotating the index $(1,2,3)$. Here we have used
the relation $\dot\phi_0^2\simeq\frac{\phi_0^4}{\lambda}$ due to
small speed of sound. %In the UV model \cite{DBI,DBI in sky}, the
%potential is of the form
%\begin{equation}
%V(\phi)\simeq\frac{1}{2}m^2\phi^2,
%\end{equation}
%with $m\gg M_{pl}/\sqrt{\lambda}$.
%0710.1812p18 also 0605045 p9
In this phase, we have power spectrum of $\phi$ perturbation \be
P_k=\frac{H^4}{4\pi^2\dot\phi_0^2}\approx
\frac{N_e^4}{4\pi^2\lambda},\ee where $N_e\sim 60$ is the e-folding
between horizon crossing and the end of inflation. According to COBE
normalization, $P_k\approx 23\times 10^{-10}$, we find
$\lambda\approx 10^{14}$ and $\dot\phi_0\approx 10^{-7}M_{pl}^2$.
Using the limit of small speed of sound,
$\phi_0\approx\dot\phi_0^{1/2}\lambda^{1/4}\approx M_{pl}$.
%If $m\sim 10^{-5}M_{pl}$ and
%$\lambda\sim 10^{14}$, we could obtain $\phi_0\sim M_{pl}$ by
%approximate estimation through the Friedmann equation \be
%3M_{pl}^2H^2=E.\ee
As a matter of result, only the first term in
(\ref{DBI}) with the following shape dominates,
\begin{eqnarray}
\mathcal{A}_1&=&
-\frac{H^4l_s^4}{32c_s^6}((k_1^a)^2(k_2^b)^2-k_1^ak_1^bk_2^ak_2^b)(\vec{k}_1\cdot\vec{k}_2)k_3^2(24\frac{1}{K^5}\nonumber\\
&&+120\frac{k_1+k_2}{K^6}+720\frac{k_1k_2}{K^7})+perm.,
\end{eqnarray}
We parameterize
$\langle\zeta_{k_1}\zeta_{k_2}\zeta_{k_3}\rangle_\theta$ as,
\be\label{parameter}
\langle\zeta_{k_1}\zeta_{k_2}\zeta_{k_3}\rangle_\theta=(2\pi)^7\delta(\vec{k}_1+\vec{k}_2+\vec{k}_3)(P_\zeta^k)^2\frac{1}{\prod_ik_i^3}\mathcal{A}_1,\ee
where $P^k_\zeta$ is the primordial power spectrum \cite{muk}
\begin{equation}
P^k_\zeta=\frac{1}{8\pi^2M_{pl}^2}\frac{H^2}{c_s\epsilon}.
\end{equation}
The shape of $\mathcal{A}_1/k_1k_2k_3$ as function of $x_1=k_1/k_3$
and $x_2=k_2/k_3$ are drawn in Fig.~\ref{A1}. In drawing the
figures, we omit the coefficient of $\mathcal{A}_1$. And we have
chosen a particular frame in which the three $\vec k$ modes are in
$x-y$ plane and $\vec k_3$ is along the $x$-axis. We find this shape
is distinct from the shape of $\mathcal{A}_c$\cite{Shiu}.

The non-Gaussianity of CMB in the WMAP observations is analyzed by
assuming the following ansatz
\begin{equation}
\zeta=\zeta_g-\frac{3}{5}f_{NL}\zeta_g^2,
\end{equation}
where $\zeta_g$ represents the Gaussian part of $\zeta$, and
$f_{NL}$ is an estimator of non-Gaussianity. The three-point
function of $\zeta$ can be factorized as
\begin{equation}\label{shape}
\langle\zeta_{\vec{k}_1}\zeta_{\vec{k}_2}\zeta_{\vec{k}_3}\rangle=(2\pi)^7\delta^3(\vec{k}_1+\vec{k}_2+\vec{k}_3)(-\frac{3}{10}f_{NL}(P^k_\zeta)^2)
\frac{\sum_ik_i^3}{\prod_ik_i^3}.
\end{equation}

Despite the difference between the shape of (\ref{DBI}) and that of
(\ref{shape}), we set $k_1=k_2=k_3=k$ to calculate $f_{NL}^1$ which
represents the size of correction of non-Gaussianity deriving from
$\mathcal{A}_1$. We have
\begin{eqnarray}
&&f_{NL}^1=0.02\frac{H^4l_s^4}{c_s^6},
\end{eqnarray}
where the results are evaluated in the particular frame we used
above.
%\begin{equation}
%\alpha=\frac{((k_1^a)^2(k_2^b)^2-k_1^ak_1^bk_2^ak_2^b)(\vec{k}_1\cdot\vec{k}_2)}{k^6}+perm..
%\end{equation}
%The range of $\alpha$ is roughly $ -1<\alpha<2$.

We find the noncommutative correction of non-Gaussianity could be
large if $c_s\ll1$. For example, for Hubble constant $H \sim
10^{-5}M_{pl}$, the string length scale $l_s\sim 10^{4}M_{pl}^{-1}$
and the speed of sound $c_s \sim 0.1$, then $f_{NL}^1\sim 2$.
Comparing to the dominant $f_{NL}$ of commutative case((5.10) in
\cite{Shiu}), \be f_{NL}^c\approx 0.32c_s^{-2},\ee we find the
correction is more sensitive to the speed of sound $c_s$. The
present observation imposes bound on equilateral form of $f_{NL}$,
$-256< f_{NL}<332$, and future observation of Planck can detect
$\vert f_{NL}\vert \gtrsim 5$, thus making the correction within the
sensibility of these observation.

\subsection{K-inflation}We now consider the correction in K-inflation\cite{muk,muk1} model which
also has small speed of sound. The Lagrangian of the power law
K-inflation is of the form
\begin{equation}
P(X,\phi)=\frac{4}{9}\frac{4-3\gamma}{\gamma^2}\frac{1}{\phi^2}(-X+X^2),
\end{equation}
where $\gamma$ is a constant. In the inflationary solution, $X$
remains constant as \be X_0=\frac{2-\gamma}{4-3\gamma}.\ee The
solution leads to scale factor $a$ of \be a\sim
t^{\frac{2}{3\gamma}}\ee for any $0<\gamma<\frac{2}{3}$. And the
speed of sound is \be\label{c} c_s^2=\frac{\gamma}{8-3\gamma}.\ee In
order to get small speed of sound, we focus on the region
$\gamma\ll1$. The power spectrum in the limit of small $\gamma$ is
\cite{muk} \be\label{P}
P^k_{\zeta}=\frac{1}{c_s}\frac{2}{3\gamma}\frac{H^2}{8\pi^2M^2_{pl}}(\frac{k}{k_1})^{-3\gamma},\ee
where $H$ is taken to be the Hubble scale at the time of horizon
exit for the perturbations currently at our horizon, and $k_1$ is
the associated comoving wavenumber. Using (\ref{c}) as well as the
fact that data determines $P^k_{\zeta}\sim 10^{-9}$ at horizon
crossing, we get \be\label{H} H^2\sim \frac{3}{4\sqrt
2}\gamma^{3/2}8\pi^2M_{pl}^2\times 10^{-9}.\ee It follows from
(\ref{H}) that the tilt satisfies \be n_s-1=-3\gamma+\cdots,\ee
which allows us to determine $\gamma\sim \frac{1}{60}$ and thus
$c_s^2\sim \frac{1}{480}$ using the central value of the spectral
index in the WMAP results\cite{WMAP3yr}.

\begin{figure}[h]%
\begin{center}%
 \mbox{\epsfxsize=8cm\epsffile{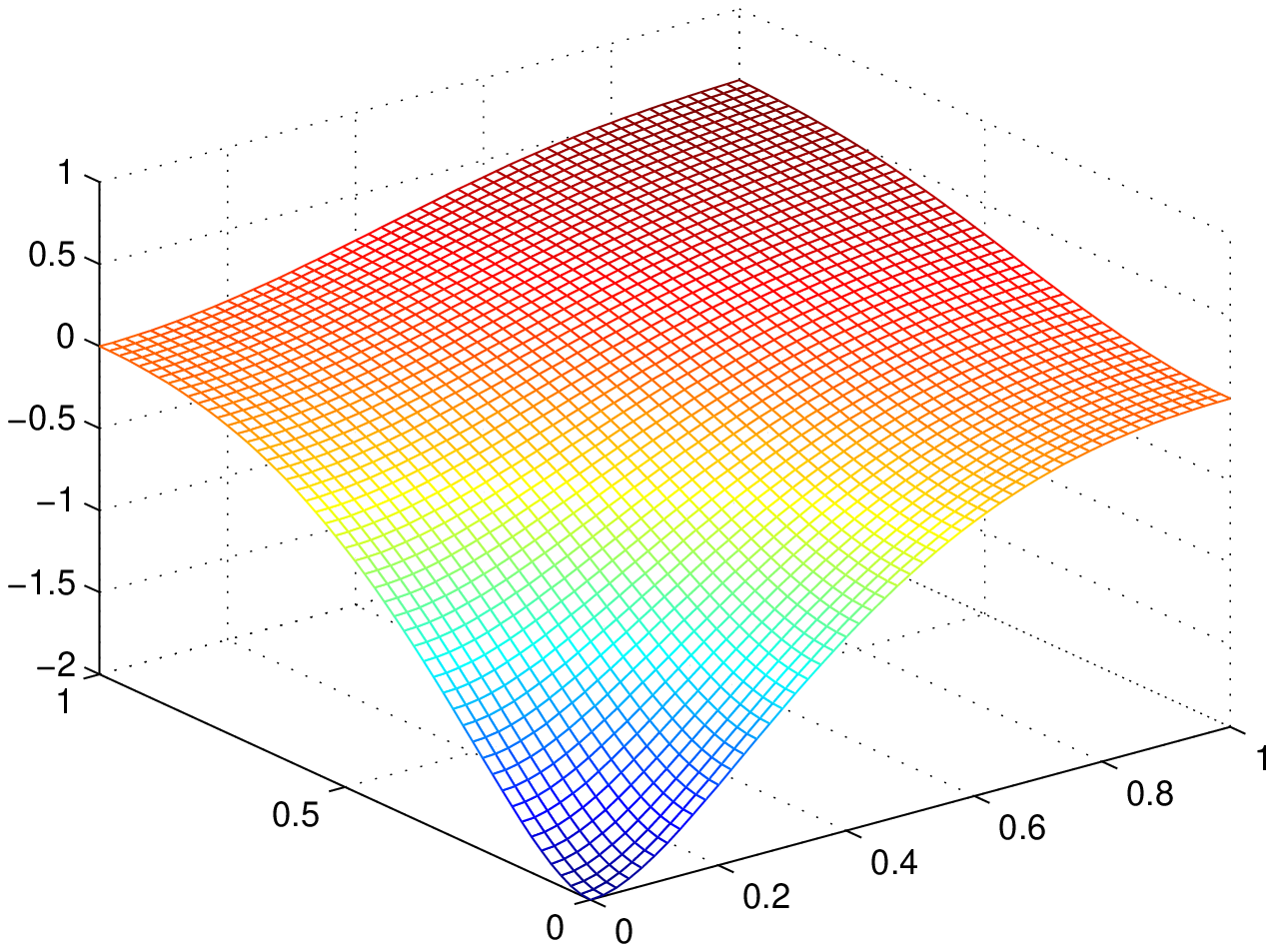}}
\caption{\label{A2}The shape of $\mathcal{A}_2/k_1k_2k_3$.}
\end{center}%
\end{figure}

\begin{figure}[h]%
\begin{center}%
\mbox{\epsfxsize=8cm\epsffile{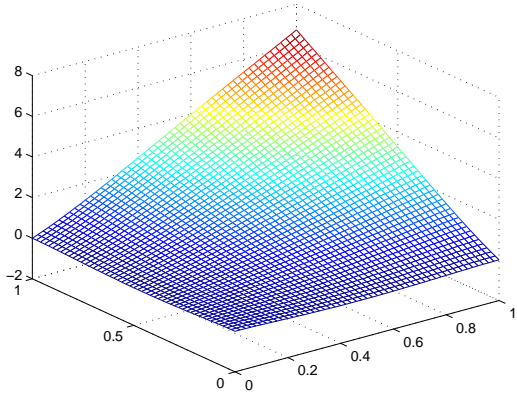}} \caption{\label{A3}The shape
of $\mathcal{A}_3/k_1k_2k_3$.}
\end{center}%
\end{figure}

We also find $c_V=-1$ and $\sigma=0$ in this model, where we choose
$V=\phi^2$. The dominant terms in three-point function are those
with most $c_s$ in the denominator in each kind of coefficients,
\begin{eqnarray}
\langle\zeta_{k_1}\zeta_{k_2}\zeta_{k_3}\rangle_\theta&=&
i(2\pi)^3\delta(\vec{k}_1+\vec{k}_2+\vec{k}_3)\frac{H^4l_s^4}{32\epsilon^2c_s^4M_{pl}^4}\frac{1}{\prod(2k_i^3)}[2i\frac{H^4}{c_s^4}
((k_1^a)^2(k_2^b)^2\nonumber\\
&&-k_1^ak_1^bk_2^ak_2^b)(\vec{k}_1\cdot\vec{k}_2)k_3^2(24\frac{1}{K^5}+120\frac{k_1+k_2}{K^6}+720\frac{k_1k_2}{K^7})\nonumber\\
&&-2\frac{iH^3}{c_s^2}\frac{M_{pl}^2}{\phi_0}((k_1^a)^2(k_2^b)^2-k_1^ak_1^bk_2^ak_2^b)(\vec{k}_1\cdot\vec{k}_2)(8\frac{1}{K^3}\nonumber\\
&&+24\frac{k_1k_2+k_2k_3+k_3k_1}{K^5}+120\frac{k_1k_2k_3}{K^6})\nonumber\\
&&+iH^2\frac{M_{pl}^4}{\phi_0^2}(((k_1^a)^2+(k_1^b)^2)(\vec{k}_2\cdot\vec{k}_3)(2\frac{1}{K}+2\frac{k_1k_2+k_2k_3+k_3k_1}{K^3}\nonumber\\
&&+6\frac{k_1k_2k_3}{K^4})+2((k_1^a)^2(k_2^b)^2-k_1^ak_1^bk_2^ak_2^b)k_3^2(2\frac{1}{K^3}\nonumber\\
&&+6\frac{k_1+k_2}{K^4}+24\frac{k_1k_2}{K^5}))+perm.].
\end{eqnarray}
Following the same parametrization as (\ref{parameter}), we get \bqa
\mathcal{A}_1&=&-\frac{H^4l_s^4}{32c_s^6}((k_1^a)^2(k_2^b)^2-k_1^ak_1^bk_2^ak_2^b)(\vec{k}_1\cdot\vec{k}_2)k_3^2(24\frac{1}{K^5}\nonumber\\
&&+120\frac{k_1+k_2}{K^6}+720\frac{k_1k_2}{K^7})+perm.,\eqa
\bqa\mathcal{A}_2&=&\frac{H^3M_{pl}^2l_s^4}{32c_s^4\phi_0}((k_1^a)^2(k_2^b)^2-k_1^ak_1^bk_2^ak_2^b)(\vec{k}_1\cdot\vec{k}_2)(8\frac{1}{K^3}\nonumber\\
&&+24\frac{k_1k_2+k_2k_3+k_3k_1}{K^5}+120\frac{k_1k_2k_3}{K^6})+perm.,\eqa
\bqa\mathcal{A}_3&=&-\frac{H^2M_{pl}^4l_s^4}{64c_s^2\phi_0^2}\big(((k_1^a)^2+(k_1^b)^2)(\vec{k}_2\cdot\vec{k}_3)(2\frac{1}{K}+2\frac{k_1k_2+k_2k_3+k_3k_1}{K^3}\nonumber\\
&&+6\frac{k_1k_2k_3}{K^4})+2((k_1^a)^2(k_2^b)^2-k_1^ak_1^bk_2^ak_2^b)k_3^2(2\frac{1}{K^3}+6\frac{k_1+k_2}{K^4}\nonumber\\
&&+24\frac{k_1k_2}{K^5})\big)+perm.. \eqa  The shape of
$\mathcal{A}_1/k_1k_2k_3$ is the same as that of DBI case. The shape
of $\mathcal{A}_2/k_1k_2k_3$ and $\mathcal{A}_3/k_1k_2k_3$ are drawn
in Fig.~\ref{A2} and Fig.~\ref{A3} respectively.

 In the equilateral triangle configuration $k_1=k_2=k_3=k$, the
correction of $f_{NL}$ is divided into $f_{NL}^1,f_{NL}^2,f_{NL}^3$,
\bqa\label{knc} f_{NL}^1=0.02\frac{H^4l_s^4}{c_s^6}, \\
f_{NL}^2=-0.02\frac{H^3l_s^4M_{pl}^2}{\phi_0c_s^4},\\
f_{NL}^3=0.25\frac{H^2l_s^4M_{pl}^4}{\phi_0^2c_s^2},\eqa

%where $\alpha$ is defined as above, and \begin{eqnarray}
%&&\beta=\frac{(k_1^a)^2+(k_1^b)^2}{k^2},\\
%&&\kappa=\frac{(k_1^a)^2(k_2^b)^2-k_1^ak_1^bk_2^ak_2^b}{k^4},
%\end{eqnarray} whose range by rough estimation are $0<\beta<1,-1<\kappa<2$.
We estimate the size of $\phi_0$ through Friedmann equation \be
3M_{pl}^2H^2=E, \ee where \be
E=2XP_{,X}-P=\frac{4}{9}\frac{4-3\gamma}{\gamma^2}\frac{1}{\phi^2}(-X+3X^2),\ee
and find $\phi_0\approx \frac{M_{pl}^2}{12\sqrt3c_s^2 H}$. Using
(\ref{c}) and (\ref{H}), we find \be
f_{NL}^2=-12\sqrt{3}c_s^4f_{NL}^1,\ee and\be
f_{NL}^3=5400c_s^8f_{NL}^1.\ee Since $c_s^2\sim \frac{1}{480}$,
these two are much smaller than $f_{NL}^1$. Using (\ref{c}),
(\ref{H}) and $l_s=10^4M_{pl}^{-1}$, we find \be
f_{NL}^1\approx180,\ee which does not depend on $c_s$ due to the
cancelation between $H^4$ and $c_s^6$. This property is different
from the case of DBI model. The dominant non-Gaussianity in
K-inflation without considering noncommutative effect is
\be\label{kng} f^c_{NL}\approx 0.26c_s^{-2}\approx 125.\ee Since
$c_s$ is fixed in K-inflation, it is easy to set an upper bound on
$l_s$. Adding (\ref{knc}) and (\ref{kng}) together, we get the total
non-gaussianity in K-inflation, \be f_{NL}^t=180\times
10^{-16}l_s^4M_{pl}^4+125.\ee Using the upper bound on the
equilateral form of $f_{NL}$, $f_{NL}<332$, we get \be l_s\leqslant
10^4M_{pl}^{-1}.\ee

\section{Conclusion}

We studied the noncommutative corrections of non-Gaussianities of
primordial perturbation in a general framework. The corrections
could be large in the models with a small speed of sound and a
relatively low string scale.  We test our result in two particular
models, DBI model and K-inflation. We find that the correction of
$f_{NL}$ can reach $O(1)$ or even bigger, and thus is observable
within the sensibility of future experiments. Our study also shows
that the noncommutative corrections are more sensitive to the
speed of sound than the usual nonGaussianity estimator. This could
be a clue to distinguish the different contributions to the
non-Gaussianity. And also it indicates that in the inflation
models with large speed of sound, the noncommutative correction to
the nonGaussianity is small. Moreover, the shape of the
corrections are different from the commutative case, which can be
used as another distinguishing estimator. %However, since the corrections are minor
%as one would expect, the shape of these correction is over-shadowed
%by the primordial shape of non-Gaussianities. We expect that by data
%analysis their impact on the shape could also be detected.

From our study, it turns out that the noncommutative corrections
become significant when the string scale is relatively low. This
could be just an illusion, given the noncommutative scale may not be
string scale. In fact, the leading order correction terms are
actually proportional to $(\theta^{12})^2$, which could be
nontrivially related to string scale. If we have a relatively low
noncommutative scale, namely we assume a little larger
noncommutativity, the correction could be larger.
 In string theory, the noncommutative scale
depends also on the background $B_{12}$ field on D-brane. In a
sense, the corrections tell us the information of the string scale
and also of the background field.

On the other hand, without assuming another scale, one can make the
natural choice (\ref{nc1}).  Although it is not easy to explain the
great difference between the Planck scale and the string scale in
the perturbative string theory, it is not hard to tune the string
length scale all the way up to $10^{-18}$cm in type I
compactifications and nonperturbative heterotic string theory. More
interestingly, in recent phenomenology study, the string scale is
fit by WMAP data to be around
$10^{-5}\sim10^{-4}M_{pl}$\cite{Huang,Tsujikawa:2003gh,LL}.
Doubtlessly, future experiment will bring forward more data to test
this tentative choice.

\section*{Acknowledgments}
The work was partially supported by NSFC Grant No.
10405028,10535060, NKBRPC (No. 2006CB805905) and the Key Grant
Project of Chinese Ministry of Education (NO. 305001). K. Fang
thanks S.-H. Henry Tye for helpful suggestion and Gary Shiu for
discussion.

\appendix

\section{ Noncommutative correction of third order perturbation} The
noncommutative correction of third order action of perturbation,
which is of order $O(\epsilon)$ except the terms proportional to
$\sigma$, is
\begin{eqnarray}
&&\delta_\theta S=\int
d^{4}xa^3\frac{1}{8}(\theta^{12})^2\Big((\frac{P_{,X_0}}{\dot\phi_0}(\frac{1}{c_s^2}-1)\dot\varphi+P_{,X_0}\frac{V_0^{'}}{2V_0}(\frac{1}{c_V^2}-1)\varphi)
(\frac{1}{a^2}\partial_1^2\partial_i\varphi\partial_2^2\partial_i\varphi\nonumber\\
&&\qquad
\qquad-\frac{1}{a^2}\partial_1\partial_2\partial_i\varphi\partial_1\partial_2\partial_i\varphi
-H\partial_1^2\partial_i\varphi\partial_i\dot\varphi-H\partial_2^2\partial_i\varphi\partial_i\dot\varphi
+H^2\partial_1^2\partial_i\varphi\partial_i\varphi+\nonumber\\
&& \qquad \qquad
H^2\partial_2^2\partial_i\varphi\partial_i\varphi+\dot
a^2\partial_1\dot\varphi\partial_1\dot\varphi+\dot
a^2\partial_2\dot\varphi\partial_2\dot\varphi+\dot
a^2\partial_i\dot\varphi\partial_i\dot\varphi   -2H\dot
a^2\partial_1\dot\varphi\partial_1\varphi\nonumber\\
&& \qquad \qquad -2H\dot
a^2\partial_2\dot\varphi\partial_2\varphi-2H\dot
a^2\partial_i\dot\varphi\partial_i\varphi+(H\dot
a)^2\partial_1\varphi\partial_1\varphi +(H\dot
a)^2\partial_2\varphi\partial_2\varphi\nonumber\\
&& \qquad \qquad +(H\dot
a)^2\partial_i\varphi\partial_i\varphi-\partial_1^2\dot\varphi\partial_2^2\dot\varphi+\partial_1\partial_2\dot\varphi\partial_1\partial_2\dot\varphi
-(Ha\dot a)^2\dot\varphi^2 -Ha\dot
a\dot\varphi\partial_1^2\dot\varphi\nonumber\\
&& \qquad \qquad-Ha\dot
a\dot\varphi\partial_2^2\dot\varphi+2H\partial_1^2\dot\varphi\partial_2^2\varphi+2H\partial_1^2\varphi\partial_2^2\dot\varphi+
2H^2a\dot a\dot\varphi\partial_1^2\varphi +2H^2a\dot
a\dot\varphi\partial_2^2\varphi\nonumber\\
&&\qquad\qquad-4H\partial_1\partial_2\dot\varphi\partial_1\partial_2\varphi-4H^2\partial_1^2\varphi\partial_2^2\varphi
+4H^2\partial_1\partial_2\varphi\partial_1\partial_2\varphi+ a \dot
a {\partial_1}^2 \dot \varphi \ddot \varphi \nonumber\\
&&\qquad \qquad+ a \dot a {\partial_2}^2 \dot \varphi \ddot \varphi
- a^2 {\dot a}^2 \ddot \varphi \ddot \varphi - {\dot a}^2 \ddot
\varphi ( 2 {\partial_1}^2 \varphi +2 {\partial_1}^2 \varphi- a \dot
a \dot
\varphi))-\frac{1}{2}\frac{P_{,X_0}}{\dot\phi_0}(\frac{1}{c_s^2}-1)Ha\dot
a\nonumber\\
&& \qquad \qquad
(\dot\varphi^2-(\partial\varphi)^2)(\partial_1^2\dot\varphi+\partial_2^2\dot\varphi-2H\partial_1^2\varphi-2H\partial_2^2\varphi+2Ha\dot
a\dot\varphi)\nonumber\\
&& \qquad
\qquad-P_{,X_0}\frac{n(n-1)\dot\phi_0}{4\phi_0^2}(\frac{1}{c_V^2}-1)(\dot\varphi^2-(\partial\varphi)^2)(2(a\dot
a)^2\dot\varphi-a\dot a\partial_1^2\varphi-a\dot
a\partial_2^2\varphi))\nonumber\\
&& \qquad
\qquad-P_{,X_0}\frac{n(n-1)\dot\phi_0}{2\phi_0^2}(\frac{1}{c_V^2}-1))(\dot\varphi\partial_1^2\varphi\partial_2^2\varphi-\dot\varphi\partial_1\partial_2\varphi\partial_1\partial_2\varphi
+(a\dot a)^2\dot\varphi^3\nonumber\\
&& \qquad \qquad-a\dot a\dot\varphi^2\partial_1^2\varphi-a\dot
a\dot\varphi^2\partial_2^2\varphi)-\frac{\sigma}{X_0^2}Ha\dot
a\dot\phi_0\dot\varphi^2(\partial_1^2\dot\varphi+\partial_2^2\dot\varphi-2H\partial_1^2\varphi-2H\partial_2^2\varphi\nonumber\\
&& \qquad \qquad+2Ha\dot a\dot\varphi)
 \Big).
\end{eqnarray}

\section{Noncommutative correction of three-point function}
The correction of three-point function of $\zeta$ is
\begin{eqnarray}\label{three}
\langle\zeta_{\vec{k}_1}\zeta_{\vec{k}_2}\zeta_{\vec{k}_3}\rangle_\theta&=&
i(2\pi)^3\delta^3(\vec{k}_1+\vec{k}_2+\vec{k}_3)\frac{H^4l_s^4}{32\epsilon^2c_s^4M_{pl}^4}\frac{1}{\prod(2k_i^3)}\Big(
2c_sH^4(\frac{1}{c_s^2}-1)\nonumber\\
&&[((k_1^a)^2(k_2^b)^2-k_1^ak_1^bk_2^ak_2^b)((\vec{k}_1\cdot\vec{k}_2)k_3^2\frac{i}{c_s^3}(24\frac{1}{K^5}+120\frac{k_1+k_2}{K^6}+720\frac{k_1k_2}{K^7})\nonumber\\
&&-720k_1^2k_2^2k_3^2\frac{i}{c_s}\frac{1}{K^7}-k_3^2\frac{i}{c_s}(8\frac{1}{K^3}+24\frac{k_1+k_2}{K^4}+96\frac{k_1k_2}{K^5}))\nonumber\\
&&+((k_1^a)^2+(k_1^b)^2)(-(\vec{k}_1\cdot\vec{k}_2)k_2^2k_3^2\frac{i}{c_s}(24\frac{1}{K^5}+120\frac{k_1}{K^6})\nonumber\\
&&+(\vec{k}_1\cdot\vec{k}_2)k_3^2\frac{i}{c_s}(2\frac{1}{K^3}+6\frac{k_1+k_2}{K^4}+24\frac{k_1k_2}{K^5})-24k_1^2k_2^2k_3^2c_s^3\frac{i}{K^5}\nonumber\\
&&+k_2^2k_3^2c_si(4\frac{1}{K^3}+12\frac{k_1}{K^4})+12ik_1^2k_2^2k_3^2c_s\frac{1}{K^5}\nonumber\\
&&+ik_2^2k_3^2c_s(2\frac{1}{K^3}+6\frac{k_1}{K^4})+k_1^2(\vec{k}_2\cdot\vec{k}_3)\frac{i}{c_s}(\frac{1}{K^3}+3\frac{k_2+k_3}{K^4}+12\frac{k_2k_3}{K^5})\nonumber\\
&&-(\vec{k}_2\cdot\vec{k}_3)\frac{i}{c_s}(\frac{2}{K}+2\frac{k_1k_2+k_2k_3+k_3k_1}{K^3}+6\frac{k_1k_2k_3}{K^4}))\nonumber\\
&&+(k_1^ak_2^a+k_1^bk_2^b+\vec{k}_1\cdot\vec{k}_2)(24ik_1^2k_2^2k_3^2c_s\frac{1}{K^5}-2ik_2^2k_3^2c_s(2\frac{1}{K^3}+6\frac{k_1}{K^4})\nonumber\\
&&+k_3^2ic_s(\frac{1}{K}+\frac{k_1+k_2}{K^2}+2\frac{k_1k_2}{K^3}))-4ic_s^3k_1^2k_2^2k_3^2\frac{1}{K^3}\nonumber\\
&&+2((k_1^a)^2(k_2^b)^2+(k_1^b)^2(k_2^a)^2)k_2^2k_3^2\frac{i}{c_s}(24\frac{1}{K^5}+120\frac{k_1}{K^6})\nonumber\\
&&-4k_1^ak_1^bk_2^ak_2^bk_2^2k_3^2\frac{i}{c_s}(24\frac{1}{K^5}+120\frac{k_1}{K^6})\nonumber\\
&&+(\vec{k}_2\cdot\vec{k}_3)k_1^2ic_s(\frac{1}{K}+\frac{k_2+k_3}{K^2}+2\frac{k_2k_3}{K^3})]\nonumber\\
&&-ik_1^2k_2^2k_3^2{c_s}[(k_1^a)^2+(k_1^b)^2](\frac{48}{K^5}+\frac{120k_2}{K^6})\nonumber\\
&&+i k_1^2k_2^2k_3^2
c_s^3(\frac{8}{K^3}-12\frac{k_1+k_2}{K^4}+24\frac{k_1
k_2}{K^5})\nonumber\\ &&+2i k_2^2k_3^2{c_s}
[(k_1^a)^2+(k_1^b)^2](\frac{4}{K^3}-6\frac{k_2-2k_1}{K^4}-24\frac{k_1k_2}{K^5})\nonumber\\
&&-2ik_1^2k_2^2k_3^2c_s^3
(\frac{4}{K^3}-6\frac{k_2}{K^4})\nonumber\\
&&+nc_sH^3(\frac{1}{c_V^2}-1)\frac{\dot\phi_0}{\phi_0}[((k_1^a)^2(k_2^b)^2-k_1^ak_1^bk_2^ak_2^b)((\vec{k}_1\cdot\vec{k}_2)\frac{i}{c_s^3}(8\frac{1}{K^3}\nonumber\\
&&+24\frac{k_1k_2+k_2k_3+k_3k_1}{K^5}+120\frac{k_1k_2k_3}{K^6})-k_1^2k_2^2\frac{i}{c_s}(24\frac{1}{K^5}+120\frac{k_3}{K^6})\nonumber
\end{eqnarray}
\begin{eqnarray}
&&\qquad\qquad\qquad-4\frac{i}{c_s}(2\frac{1}{K}+2\frac{k_1k_2+k_2k_3+k_3k_1}{K^3}+6\frac{k_1k_2k_3}{K^4}))\nonumber\\
&&\qquad\qquad\qquad+((k_1^a)^2+(k_1^b)^2)(-(\vec{k}_1\cdot\vec{k}_2)k_2^2\frac{i}{c_s}(2\frac{1}{K^3}+6\frac{k_1+k_3}{K^4}+24\frac{k_1k_3}{K^5})\nonumber\\
&&\qquad\qquad\qquad+(\vec{k}_1\cdot\vec{k}_2)\frac{i}{c_s}(2\frac{1}{K}+2\frac{k_1k_2+k_2k_3+k_3k_1}{K^3}+6\frac{k_1k_2k_3}{K^4})\nonumber\\
&&\qquad\qquad\qquad-k_1^2k_2^2ic_s(2\frac{1}{K^3}+6\frac{k_3}{K^4})+2k_2^2ic_s(\frac{1}{K}+\frac{k_1+k_3}{K^2}+2\frac{k_1k_3}{K^3}))\nonumber\\
&&\qquad\qquad\qquad-(k_1^ak_2^a+k_1^bk_2^b+\vec{k}_1\cdot\vec{k}_2)(-k_1^2k_2^2ic_s(2\frac{1}{K^3}+6\frac{k_3}{K^4})\nonumber\\
&&\qquad\qquad\qquad+2k_1^2ic_s(\frac{1}{K}+\frac{k_2+k_3}{K^2}+2\frac{k_2k_3}{K^2})\nonumber\\
&&\qquad\qquad\qquad-ic_s(K+\frac{k_1k_2+k_2k_3+k_3k_1}{K}+\frac{k_1k_2k_3}{K^2}))-k_2^2k_3^2ic_s^3(\frac{1}{K}+\frac{k_1}{K^2})\nonumber\\
&&\qquad\qquad\qquad+2((k_1^a)^2(k_2^b)^2+(k_1^b)^2(k_2^a)^2)k_1^2\frac{i}{c_s}(2\frac{1}{K^3}+6\frac{k_2+k_3}{K^4}+24\frac{k_2k_3}{K^5})\nonumber\\
&&\qquad\qquad\qquad-4k_1^ak_1^bk_2^ak_2^bk_1^2\frac{i}{c_s}(2\frac{1}{K^3}+6\frac{k_2+k_3}{K^4}+24\frac{k_2k_3}{K^5})\nonumber\\
&&\qquad\qquad\qquad-ik_1^2k_2^2 {c_s}
((k_1^a)^2+(k_1^b)^2)(\frac{4}{K^3}-6\frac{k_2-2k_3}{K^4}-24\frac{k_2
k_3}{K^5})\nonumber\\
&&\qquad\qquad\qquad-ik_1^2k_2^2
c_s^3(\frac{4}{K}-\frac{k_1+k_2-2k_3}{K^2}+2\frac{k_1k_2-2k_2k_3-2k_1k_3}{K^3}+6\frac{k_1k_2k_3}{K^4})\nonumber\\
&&\qquad\qquad\qquad+2ik_2^2 {c_s}((k_1^a)^2+(k_1^b)^2)
(\frac{2}{K}-\frac{k_2-2k_1-2k_3}{K^2}-2\frac{k_1k_2+k_2k_3-2k_1k_3}{K^3}\nonumber\\
&&\qquad\qquad\qquad+6\frac{k_1k_2k_3}{K^4})-2i k_1^2k_2^2 c_s^3
(\frac{2}{K}-\frac{k_2-2k_3}{K^2}+\frac{2k_2k_3}{K^3})]\nonumber\\
&&\qquad\qquad\qquad-\frac{n(n-1)}{2}c_sH^2(\frac{1}{c_V^2}-1)\frac{\dot\phi_0^2}{\phi_0^2}[8k_1^2k_2^2k_3^2ic_s^3\frac{1}{K^3}\nonumber\\
&&\qquad\qquad\qquad+((k_1^a)^2+(k_1^b)^2)(-3k_2^2k_3^2ic_s(2\frac{1}{K^3}+6\frac{k_1}{K^4})\nonumber\\
&&\qquad\qquad\qquad+\vec{k}_2\cdot\vec{k}_3\frac{i}{c_s}(2\frac{1}{K}+2\frac{k_1k_2+k_2k_3+k_3k_1}{K^3}+6\frac{k_1k_2k_3}{K^4}))\nonumber\\
&&\qquad\qquad\qquad-2(\vec{k}_1\cdot\vec{k}_2)k_3^2ic_s(\frac{1}{K}+\frac{k_1+k_2}{K^2}+2\frac{k_1k_2}{K^3})\nonumber\\
&&\qquad\qquad\qquad+2((k_1^a)^2(k_2^b)^2-k_1^ak_1^bk_2^ak_2^b)k_3^2\frac{i}{c_s}(2\frac{1}{K^3}+6\frac{k_1+k_2}{K^4}+24\frac{k_1k_2}{K^5})]\nonumber\\
&&\qquad\qquad\qquad-\frac{4\sigma c_sH^2}{M_{pl}^2\epsilon}[((k_1^a)^2+(k_1^b)^2)(24k_1^2k_2^2k_3^2ic_s\frac{1}{K^5}\nonumber\\
&&\qquad\qquad\qquad-2k_2^2k_3^2ic_s(2\frac{1}{K^3}+6\frac{k_1}{K^4})+4k_1^2k_2^2k_3^2ic_s^3\frac{1}{K^3})]+perm.\Big),
\end{eqnarray}
where $a$ and $b$ denote the first and second component of $k$
vector respectively, $K=k_1+k_2+k_3$ and $perm.$ denotes all the
other terms obtained by rotating the index $(1,2,3)$. All the
background value and Hubble constant are calculated at horizon
crossing.

\end{document}